\documentclass[sigconf]{acmart}

\AtBeginDocument{%
  }

\setcopyright{acmlicensed}
\copyrightyear{2025}
\acmYear{2025}
\acmDOI{XXXXXXX}

\acmConference[WWW'25]{Proceedings of the ACM
Web Conference 2025}{28 April - 2 May, 2025}{Sydney, Australia}
\acmISBN{978-1-4503-XXXX-X/18/06}

\begin{document}

\title{A Tutorial of Personalized Federated Recommender Systems: Recent Advances and Future Directions}

\author{Jing Jiang}
\affiliation{%
  \institution{Australian Artificial Intelligence Institute, University of Technology Sydney}
  \city{Sydney}
  \country{Australia}
}
\email{jing.jiang@uts.edu.au}

\author{Chunxu Zhang}
\affiliation{%
  \institution{College of Computer Science and Technology, Jilin University}
  \city{Jilin}
  \country{China}}
\email{zhangchunxu@jlu.edu.cn}

\author{Honglei Zhang}
\affiliation{%
  \institution{School of Computer Science and Technology, Beijing Jiaotong University}
  \city{Beijing}
  \country{China}
}
\email{honglei.zhang@bjtu.edu.cn}

\author{Zhiwei Li}
\affiliation{%
 \institution{Australian Artificial Intelligence Institute, University of Technology Sydney}
 \city{Sydney}
  \country{Australia}
 }
\email{zhiwei.li@student.uts.edu.au}

\author{Yidong Li}
\affiliation{%
  \institution{School of Computer Science and Technology, Beijing Jiaotong University}
  \city{Beijing}
  \country{China}
  }
\email{ydli@bjtu.edu.cn}

\author{Bo Yang}
\affiliation{%
  \institution{College of Computer Science and Technology, Jilin University}
  \city{Jilin}
  \country{China}}
\email{ybo@jlu.edu.cn}

\renewcommand{\shortauthors}{Jing Jiang et al.}

\begin{abstract}

  Personalization stands as the cornerstone of recommender systems (RecSys), striving to sift out redundant information and offer tailor-made services for users. However, the conventional cloud-based RecSys necessitates centralized data collection, posing significant risks of user privacy breaches. In response to this challenge, federated recommender systems (FedRecSys) have emerged, garnering considerable attention. FedRecSys enable users to retain personal data locally and solely share model parameters with low privacy sensitivity for global model training, significantly bolstering the system's privacy protection capabilities. Within the distributed learning framework, the pronounced non-iid nature of user behavior data introduces fresh hurdles to federated optimization. Meanwhile, the ability of federated learning to concurrently learn multiple models presents an opportunity for personalized user modeling. Consequently, the development of personalized FedRecSys (PFedRecSys) is crucial and holds substantial significance. This tutorial seeks to provide an introduction to PFedRecSys, encompassing (1) an overview of existing studies on PFedRecSys, (2) a comprehensive taxonomy of PFedRecSys spanning four pivotal research directions—client-side adaptation, server-side aggregation, communication efficiency, privacy and protection, and (3) exploration of open challenges and promising future directions in PFedRecSys. This tutorial aims to establish a robust foundation and spark new perspectives for subsequent exploration and practical implementations in the evolving realm of RecSys.
\end{abstract}




\maketitle

\section{Topic and Relevance}
As a crucial tool in combating information overload on the web, recommender systems (RecSys) are designed to leverage user behavior data to filter out irrelevant items and provide tailored services. RecSys play a pivotal role across a spectrum of domains, notably in e-commerce, aiding users in finding items customized to their preferences, thereby driving sales and enhancing customer satisfaction. Their significance and influence are extensively documented in academic research and industry analyses, showcasing their capacity to generate economic value and elevate user contentment.

Traditional cloud-based RecSys often rely on centralized data collection, raising significant concerns regarding user privacy and data security compliance, such as GDPR regulations. To address these challenges, federated recommender systems (FedRecSys) have emerged and gained prominence. FedRecSys alow users to store data locally and share only essential model parameters for global training. This decentralized approach not only enhances privacy protection and regulatory compliance but also fosters user trust, presenting a promising alternative to centralized systems.


However, in contrast to conventional federated learning scenarios involving a limited number of clients, FedRecSys encompasses a vast array of clients (users) with highly diverse data distributions, thus presenting formidable challenges rooted in non-iid data.  Moreover, the distributed optimization approach in federated learning enables the system to learn distinct models for clients, compelling us to design personalization modeling techniquez for FedRecSys. These dual factors drive the imperative need to establish personalized FedRecSys (PFedRecSys) to enhance user modeling and deliver superior recommendation services. From the perspectives of personalization modeling mechanisms and system optimization challenges, PFedRecSys can be categorized into four key areas: (1) client-side adaptation, which focuses on locally adapting the global model to accommodate user personalization; (2) server-side aggregation, involving the design of innovative aggregation methods to learn customized model parameters for clients; (3) communication efficiency, aiming to reduce the communication cost of transmitting model parameters between clients and the server; and (4) privacy and protection enhancements to fortify the system's privacy protection capabilities. These research directions are pivotal in shaping the future of PFedRecSys and propelling the field towards more personalized and secure recommendation experiences.

\subsection{Relevance and References}
\textbf{Relevance to the Web Conference (WWW).}
The International World Wide Web Conference (WWW) is a prestigious academic conference that holds significant influence in the field of web-related research, including the domain of RecSys. Over the past years, WWW has featured numerous papers that focus on the theme of RecSys, reflecting their importance and the growing interest from both academia and industry. In particular, the topics requested for submission by the WWW conference include a range of themes concerning RecSys, security and privacy, as well as federated learning. For instance, over the past four years, WWW conferences have featured tracks focusing on "User Modeling". Notably, the WWW'25 call for papers specifically highlights the "Federated recommendation systems and personalization" subtopic within the "User modeling, personalization, and recommendation" track. This alignment underscores the significant relevance of our tutorial on PFedRecSys topic to the evolving trends in the field, emphasizing its necessity and strong fit with the themes of the WWW conference.

With WWW'25 being a hub for leading researchers and industry experts in the PFedRecSys domain, it stands as a promising platform to share foundational insights, showcase recent research findings, and cultivate partnerships to elevate PFedRecSys. Drawing upon the wealth of knowledge and collaborative spirit of attendees, this tutorial aims to propel the field forward and tackle the inherent challenges linked to PFedRecSys technologies, fostering innovation and advancement in the realm of tailored user experiences.

\textbf{References to Related Tutorials.}
We conduct a thorough survey for the related tutorials in the past five years at the top-tier conferences and list them as follows,

\begin{itemize}
    \item Hongzhi Yin, Tong Chen, Liang Qu and Bin Cui. "On-Device Recommender Systems: A Tutorial on The New-Generation Recommendation Paradigm", in WWW'24.
    \item Wenqi Fan, Xiangyu Zhao, et al. "Trustworthy Recommender Systems: Foundations and Frontiers", in KDD'23.
    \item Wenqi Fan, Xiangyu Zhao, et al. "Trustworthy Recommender Systems: Foundations and Frontiers", in IJCAI'23
    \item Wenqi Fan, Xiangyu Zhao, et al. "Trustworthy Recommender Systems", in WWW'23
    \item Qiang Yang, Ben Tan. "Federated Recommender Systems", in IJCAI'20.
\end{itemize}

Among the related tutorials, only the one on "Federated Recommender System" aligns closely with our tutorial, while others touch upon distributed optimization methods or privacy and security aspects, only sharing some intersections with our tutorial. The "Federated Recommender System" tutorial categorizes FedRecSys broadly into horizontal, vertical, and transfer learning perspectives, focusing on corresponding implementation with open-source tool FATE. In contrast, our tutorial delves into personalized modeling perspective within FedRecSys, summarizing research achievements in personalized techniques and exploring promising future research directions. This tutorial, or any variants thereof, has not been featured at other conferences. By harnessing the insights and diverse expertise of attendees, this tutorial aims to propel the field forward and confront the existing hurdles linked to PFedRecSys studies.

\subsection{Tutorial Team}
\subsubsection{Organizer Details}

\begin{itemize}
    \item 
    \textbf{Prof. Jing Jiang}
    is currently an Associate Professor in the School of Computer Science and a core member of the Australian Artificial Intelligence Institute (AAII), at the University of Technology Sydney (UTS). She is an ARC DECRA Fellow. Over the past six years, she has been granted a total of funds for more than AU\$2 million, including two ARC DPs, one ARC LP, one CISRO/Data61 CRP project (lead CI), and seven industry-funded research projects. Her research interests focus on machine learning and its applications. She has published over 80 papers in the related areas of AI in top-tier conferences and journals, such as NeurIPS, ICML, ICLR, AAAI, IJCAI, KDD and IEEE TPAMI, TNNLS, and TKDE. Dr. Jiang has been invited to be a (senior) member of the program committee for many top-tier conferences, including ICML, NeurIPS, ICLR, AAAI, IJCAI, AAMAS, and CVPR, and regularly served as a reviewer for highly rated journals, such as IEEE TPAMI, TNNLS, and TKDE. She will serve as the local co-chair for Webconf 2025. She served as program co-chair for ADMA2023, local co-chair for ADMA2021, AI2021, AusDM2015, and financial chair for ICDS2015. She was an awardee of the Australian International Postgraduate Research Scholarship (IPRS).

    \item 
    \textbf{Dr. Chunxu Zhang} 
    is currently a Postdoctoral Fellow at Jilin University, specializing in federated learning and recommender systems. She has authored over 20 papers in prestigious venues, including WWW, KDD, SIGIR, IJCAI, AAAI, and NeurIPS. Additionally, she has served as a reviewer for numerous renowned conferences and journals, such as WWW, KDD, ICLR, IJCAI, TNNLS, and TIFS.

    \item \textbf{Mr. Honglei Zhang} is currently pursuing his Ph.D. at Beijing Jiaotong University. During this period, he was a visiting scholar at Nanyang Technological University. His research work has been published on top-tier international venues, including ICDE, TOIS, and TNNLS. Additionally, he has been an PC and/or reviewer for many leading venues, such as WWW, NeurIPS, ICLR, TKDE and TOIS. His open-source tutorial RSPapers/RSAlgorithms on GitHub has garnered about 7,000 stars. He also served as the publicity chair for the Responsible Recommender workshop at KDD’22. His research interest primarily focuses on developing efficient and privacy-preserving recommender systems, with a particular interest in federated recommendation.

    \item 
    \textbf{Mr. Zhiwei Li} currently is pursuing his PhD at the University of Technology Sydney. 
    His research focuses on advancing personalized federated collaborative filtering by developing models that improve recommendation accuracy while ensuring user privacy. 
    He has authored several influential papers, contributing to key areas of federated recommendation, including personalized item representations, multimodal data integration, and the application of foundation models in federated environments. 
    Zhiwei also serves as a reviewer for leading conferences such as NeurIPS, ICLR, WWW and AISTATS, demonstrating his expertise.

    \item 
    \textbf{Prof. Yidong Li} is currently a Professor, Executive Dean of the School of Computer Science and Technology at Beijing Jiaotong University and Director of the Key Laboratory of Big Data and Artificial Intelligence in Transportation, Ministry of Education. His research interests include data security and privacy protection, advanced computing and intelligent recommender systems. He has published over 200 papers in major international journals, including TIFS, TKDE, TOIS, TNNLS, TSC, and TMC, as well as at conferences such as NeurIPS, SIGKDD, CVPR, IJCAI and AAAI.

    \item 
    \textbf{Prof. Bo Yang}
     is currently the Dean of the College of Computer Science and Technology, Jilin University, Changchun, China, where he is also the Director of the Key Laboratory of Symbolic Computation and Knowledge Engineering, Ministry of Education. His research interests encompass data mining, complex network analysis, neural-symbolic systems, intelligent recommendation systems, among others. With a publication record of approximately 200 papers in esteemed journals and conferences like IEEE TPAMI, IEEE TKDE, IEEE TCYB, IEEE TNNLS, NeurIPS, ICLR, ICML, WWW, KDD, ACL, AAAI, and IJCAI.
    
\end{itemize}

\subsubsection{Relevant Publications by Organizers} To demonstrate the presenters' expertise in the field of PFedRecSys, we will present a list of papers authored by the presenters on this topic.

\begin{itemize}
    \item Client-side Adaptation~\cite{li2023federated,li2024feddae,li2024personalized,zhang2023dual,zhang2024gpfedrec,zhang2024federated,lightfr_2022,zhang2024transfr}
    \item Server-side Aggregation~\cite{attention_2019,privfr_2024,fedca_2024,zhang2024gpfedrec}
    \item Communication Efficiency~\cite{lightfr_2022,li2023federated,fedpa_2024,privfr_2024}
    \item Privacy and Protection~\cite{lightfr_2022,li2023federated,zhang2024federated}
\end{itemize}

\section{Tutorial Style}
This tutorial, designed in a lecture-style format, seeks to offer an in-depth overview of research on PFedRecSys, covering cutting-edge research and investigating future research directions.

\section{Schedule}
The tutorial spans 3 hours and comprises 5 distinct sections. Below, we present a breakdown of the tutorial's outline.

\noindent \textbf{Section 1. Welcome \& Introduction (10 mins)}

\begin{itemize}
    \item Overview of FedRecSys
    \item PFedRecSys: Background and Applications
\end{itemize}

\noindent \textbf{Section 2. Definition \& Taxonomy of PFedRecSys (20 mins)}

\begin{itemize}
    \item Definition of PFedRecSys
    \item Taxonomy of PFedRecSys
\end{itemize}

\noindent \textbf{Section 3. A Review of PFedRecSys (110 mins)}

3.1 Client-side Adaptation
\begin{itemize}
      \item User Embedding Personalization \cite{wu2021hierarchical,perfedrec_2022,lightfr_2022,li2023federated,zhang2023dual,yuan2024hetefedrec,zhang2024gpfedrec,zhang2024federated}
      
      \item Item Embedding Personalization \cite{li2023federated,zhang2023dual,li2024personalized,zhang2024gpfedrec,zhang2024federated}
        
      \item Network Parameter Personalization \cite{li2024feddae,li2024personalized,yan2024federated,zhang2024federated,zhang2024transfr}
\end{itemize}

3.2 Server-side Aggregation

\begin{itemize}
    \item Clustering-based Aggregation~\cite{pfa_2021,ifca_2020,co_cluster_2024,perfedrec_2022}
    \item Attention-based Aggregation~\cite{attention_2019,layer_attention_2022,elastic_attention_2023}
    \item Graph-based Aggregation~\cite{pfedgraph_2023,fedsac_2024,fedgkd_2023,zhang2024gpfedrec}
    \item Composite Aggregation~\cite{fedfast_2020,yuan2024hetefedrec,privfr_2024,fedca_2024,fedaf_2024}
\end{itemize}

3.3 Communication Efficiency

\begin{itemize}
    \item Reinforcement Learning-based Methods~\cite{khan2021payload,ali2024communication,di2024fedrl}
    \item Knowledge Distillation-based Methods~\cite{wang2020next,wu2022communication,zhang2024transfr,xia2022device}
    \item Hashing-based Methods~\cite{xia2023efficient,lightfr_2022,privfr_2024,yang2024discrete}
    \item Low-rank Factorization-based Methods~\cite{li2023federated,colr_2024,fedloca_2024,fedpa_2024}
    \item Neural Architecture Search-based Methods~\cite{pan2021privacy,chen2021learning,zheng2024personalized}
\end{itemize}

3.4 Privacy and Protection

\begin{itemize}
  \item Privacy Leakage \cite{chai2020secure,chai2022efficient,ding2023combining,li2023federated,yi2023ua,yu2023untargeted,yuan2023manipulating,zhang2024comprehensive,zhang2023new}
  \item Protection Measures \cite{chai2020secure,wang2020federated,wu2021fedgnn,lin2020fedrec,liang2021fedrec++,lin2021fr,zhang2021vertical,chai2022efficient,liu2022federated,li2023federated,zhang2024federated}
\end{itemize}

\noindent \textbf{Section 4. Open Challenges \& Future Directions (20 mins)}

\begin{itemize}
    \item Open Challenges in PFedRecSys
    \item Emerging Future Directions of PFedRecSys
\end{itemize}

\noindent \textbf{Section 5. Open Discussion Session (20 mins)}


\section{Audience}
Our tutorial proposal is meticulously tailored for the conference’s tech-savvy audience, with a strong emphasis on PFedRecSys. It caters to a diverse audience, from newcomers in RecSys to seasoned experts in federated learning, providing a comprehensive dive into personalized recommendations within federated settings. By intricately examining the nuances of personalized user modeling, addressing data privacy concerns, and navigating technical challenges in federated systems, our tutorial equips participants with the knowledge to expertly navigate this intricate landscape. This session highlights the latest trends and fosters collaborative discussions, enriching the conference with its focus on this cutting-edge field. Positioned to significantly enhance the value of WWW’25, our tutorial shines a spotlight on a critical and burgeoning area in RecSys. It serves as a platform for learning, networking, and contributing to industry advancements, embodying a cornerstone for knowledge exchange and professional growth.


\section{Tutorial Materials}
Upon the tutorial's approval, participants will have access to the slides and video teaser a week before the conference, enhancing their readiness for the event.

\section{Video Teaser}
The video teaser is available at \url{https://bit.ly/pfedrecsys}.


\bibliographystyle{ACM-Reference-Format}
\bibliography{sample-base}

\end{document}